# Structure determination and coexistence of superconductivity and antiferromagnetic order in $(Li_{0.8}Fe_{0.2})OHFeSe$

X. F. Lu[1], N. Z. Wang[1], H. Wu[2,6], Y. P. Wu[1], D. Zhao[1], X. Z. Zeng[1], X. G. Luo[1], T. Wu[1], W. Bao[3], G. H. Zhang[4,5], F. Q. Huang[4,5], Q. Z. Huang[2], and X. H. Chen[1]

1. *Hefei National Laboratory for Physical Sciences at Microscale and Department of Physics, University of Science and Technology of China, Hefei, Anhui 230026, China*

2. *National Institute of Standards and Technology, Center for Neutron Research, 100 Bureau Dr., Gaithersburg, Maryland 20878, USA*

3. *Department of Physics, Renmin University of China, Beijing 100872, China*

4. *CAS Key Laboratory of Materials for Energy Conversion, Shanghai Institute of Ceramics, Chinese Academy of Sciences, Shanghai 200050, China*

5. *Beijing National Laboratory for Molecular Sciences and State Key Laboratory of Rare Earth Materials Chemistry and Applications, College of Chemistry and Molecular Engineering, Peking University, Beijing 100871, China*

6. *Department of Materials Science and Engineering, University of Maryland, College Park, Maryland 20742, USA*

FeSe-derived superconductors show some unique behaviors relative to iron-pnictide superconductors, which are very helpful to understand the mechanism of superconductivity in high-$T_c$ iron-based superconductors. The low-energy electronic structure of the heavily electron-doped $A_xFe_2Se_2$ (A=K, Rb, Cs) demonstrates that interband scattering or Fermi surface nesting is not a necessary ingredient for the unconventional superconductivity in iron-based superconductors[1,2]. The superconducting transition temperature ($T_c$) in the one-unit-cell FeSe on $SrTiO_3$ substrate can reach as high as ~65 K[3-6], largely transcending the bulk $T_c$ of all known iron-based superconductors. However, in the case of $A_xFe_2Se_2$, the inter-grown antiferromagnetic insulating phase makes it difficult to study the underlying physics. Superconductors of alkali metal ions and $NH_3$ molecules or organic-molecules intercalated FeSe[7-9] and single layer or thin film FeSe on $SrTiO_3$ substrate [3-6] are extremely air-sensitive, which prevents the further investigation of their physical properties. Therefore, it is urgent to find a stable and accessible FeSe-derived superconductor for physical property measurements so as to study the underlying mechanism of superconductivity. Here, we report the air-stable superconductor $(Li_{0.8}Fe_{0.2})OHFeSe$ with high temperature superconductivity at ~40 K synthesized by a novel hydrothermal method. The crystal structure is unambiguously determined by the combination of X-ray and neutron powder diffraction and nuclear magnetic resonance. It is also found that an antiferromagnetic order coexists with superconductivity in such new FeSe-derived superconductor. This novel synthetic route opens a new avenue for exploring other superconductors in the related systems. The combination of different structure characterization techniques helps to complementarily determine and understand the details of the complicated structures

The tetragonal β-Fe$_{1+\delta}$Se (0.01≤δ≤0.04) exhibits superconductivity at ~8.5K when δ=0.01, which differs from canonical iron-based superconductors in which superconductivity arises when they closely approach the chemical stoichiometry[10,11]. In addition, unlike LnFeAsO[12,13] or BaFe$_2$As$_2$[14,15] which needs chemical substitution to drive the system from itinerant antiferromagnetic state into superconducting state, undoped FeSe exhibits superconductivity without any magnetic ordering[10,11] and T$_c$ would be greatly increased to 37 K at 7 GPa via hydrostatic pressure[16,17]. All of the cases mentioned above suggest that FeSe-derived superconductors may be good candidates to investigate the mechanism of high-T$_c$ superconductivity in the iron-based superconductors. In iron-selenides, alkali metal ions and small molecules could be intercalated into two adjacent FeSe layers to form high-Tc superconductors, including A$_x$Fe$_{2-y}$Se$_2$ (A = K, Rb, Cs, etc.) [18-21], Li$_x$(NH$_2$)$_y$(NH$_3$)$_{1-y}$Fe$_2$Se$_2$[8] and Li$_x$(C$_5$H$_5$N)$_y$Fe$_{2-z}$Se$_2$[9]. In the case of K$_x$Fe$_{2-y}$Se$_2$, the superconducting phase without Fe vacancy is always inter-grown with an insulating antiferromagnetic ordered phase which is named as K$_2$Fe$_4$Se$_5$ (245 phase) with Néel temperature as high as ~560 K and a $\sqrt{5} \times \sqrt{5}$ vacancy order of Fe[22-24]. The intergrowth of superconducting phase and the insulating 245 phase makes it difficult to investigate the underlying physics in this system. Moreover, for the compounds with alkali ions and small molecules intercalation between FeSe layers[17,18], the complexity of structure brought by the spacer layer would further hinder the investigation of the mechanism of superconductivity. Finally, the extreme air-sensitivity of these superconductors prevents the feasible physical property measurements. Therefore, it is crucial to develop suitable FeSe-derived superconductors for physical property measurements in order to reveal the underlying mechanism of superconductivity among this class of intriguing compounds. Herein, we report the comprehensive determination of the crystal structure of a novel superconductor $(Li_{0.8}Fe_{0.2})OHFeSe$ with coexistence of superconductivity and antiferromagnetic order.

The crystal structure of the sample in the present study has been determined by X-ray powder compound and adopts a structure with alternating layers of anti-PbO-type FeSe and anti-PbO-type $LiFeO_2$. All Bragg peaks of the sample can be indexed by a tetragonal structure with space group *P*4/*nmm* (No. 129), and fitted with the good *R* factors (e.g. $R_{wp}$ = 0.0937, $R_p$=0.0656). The obtained lattice parameters of $LiFeO_2Fe_2Se_2$ are a = 3.7926(1) Å, c = 9.2845(1) Å at room temperature. The lattice parameters *a* and *b* are very close to that (3.7734(1) Å) of $\beta\text{-}Fe_{1.01}Se$, while the *c*-axis is expanded by 68% compared to $\beta\text{-}Fe_{1.01}Se$ due to the intercalation of $LiFeO_2$ layer as a spacer layer into two adjacent FeSe layers[25]. However, the distance between O and Se, 3.6028(1) Å, is rather large while X-ray diffraction could not provide further information. To understand why the space between O and Se is so large and to reveal the mystery may hide between the two anions, subsequent neutron powder diffraction (NPD) experiments were conducted with Ge311 (λ=2.0775 Å) and Cu311 (λ=1.5401 Å) monochrometers using BT-1 powder diffractometer at NIST Center for Neutron Research (NCNR). Data were collected in the 2-Theta range of 1.3-166.3° (Ge311) and 3-168° (Cu311) with a step size of 0.05° in the temperature range from 2.5 K to 295 K. Unfortunately, the structure model proposed based on XRD data for $LiFeO_2Fe_2Se_2$ cannot fit the observed NPD pattern well. The neutron scattering Fourier difference map on the observed patterns using the XRD proposed structure model indicated the large neutron scattering difference near the position (0.75, 0.75, 0.178), as shown in pink color contours in Fig.1(a). The distance between the mass center of this Fourier difference contour and the O position is about 0.95 Å, which is very close to the generally reported O-H single bond (~0.96 Å, or in the range of 0.92-0.98 Å). Also the relatively large background of the NPD pattern indicates the large incoherent scattering from the possible H in the sample. Therefore, H was considered in the new structural model with its position suggested by Fourier difference analysis, while the H cannot be detected from XRD method. However, Analysis using NPD data indicates that no atoms are presence at the 2*a* site where should be occupied by Li/Fe proposed by analyzing the XRD data.

The main question to discern the difference between the structure models proposed by XRD and NPD is whether Li and H atoms exist in the crystal structure or not. Therefore, we performed site-selected nuclear magnetic resonance (NMR) experiment on the same samples because the NMR experiments can detect both Li and H elements, which is very suitable to solve the question. In this case, we used $^7$Li or $^1$H NMR to determine which nucleus is contained in structure. Surprisingly, both $^7$Li and $^1$H signals were observed and show a similar multi-peaks feature at low temperature in our samples as shown in Fig.2. The special multi-peaks structure in NMR spectra arises from the low-temperature magnetic ordering phase confirmed by susceptibility and specific heat measurements later on. As the temperature increases, both $^7$Li and $^1$H spectra become narrow and the multi-peaks feature in the spectra disappears. On the other hand, temperature-dependent spin-spin relaxation ($T_2$) for both nuclei also shows a similar divergent behavior with decreasing temperature, consistent with a magnetic phase transition. If $^7$Li and $^1$H signals are from different crystal structures not in the same crystal structure, different behaviors are expected for them. Therefore, the NMR result definitely proves that both of Li and H should be contained in the structure.

Based on the fact of both Li and H in the same crystal structure from NMR measurements and random occupation of Li/Fe at 2*a* sites, we can understand the distinct structure models proposed by XRD and NPD. The scattering factor of the lightest H element is too small to be detected by XRD. While the total neutron scattering from the 2*a* site could be 0 when the refined fraction of Li and Fe on that site is to 0.81(1) and 0.19(1), respectively, (the neutron scattering amplitudes are -2.03 fm for Li and 9.54 fm for Fe). Using the new structure model containing both of Li and H, namely "$(Li_{0.8}Fe_{0.2})(OH)(FeSe)$" (see Fig.1(b)), the 2.5 K to 295 K NPD patterns can be perfectly fitted with excellent goodness of fit factors. A plot of the observed and calculated intensities measured at 2.5 K using Ge311 monochrometer ($\lambda$=2.0775 Å) is shown in Fig.1(c) and the refined structure parameters are listed in Table I.

In the new structure model, O-H bond distance is 0.94 Å, while the H-Se distance is 3.078 Å at 295 K. Compared with 2.75 Å of H-Se distance observed in the $Li_{0.6}(ND_2)_{0.2}(ND_3)_{0.8}Fe_2Se_2$[17], the much larger H-Se distance in $(Li_{0.8}Fe_{0.2})OHFeSe$ suggests a much weaker hydrogen bonding interaction between $(Li_{0.8}Fe_{0.2})(OH)$ layers and FeSe layers. Moreover, the Se-Fe-Se bond angles are 103.2(2) °(x2) in $(Li_{0.8}Fe_{0.2})OHFeSe$, which is smaller than 103.9 °(x2) in $\beta$-$Fe_{1+\delta}Se$. Besides, the Fe-Se bond distance in $(Li_{0.8}Fe_{0.2})OHFeSe$, 2.416 Å, is larger than 2.3958 Å of $\beta$-$Fe_{1+\delta}Se$. All these structural changes indicate that the $(Li_{0.8}Fe_{0.2})OHFeSe$ owns a very distorted $FeSe_4$ tetrahedron which is compressed in the *ab* plane. Such distorted $FeSe_4$ tetrahedron has also been observed in $Li_{0.6}(NH_2)_{0.2}(NH_3)_{0.8}Fe_2Se_2$ with a $T_c$ of 43 K, which implies that a tetrahedron distortion in FeSe-derived superconductors may promote the superconductivity. Finally, different from $Fe_{1+\delta}Se$ there is no crystal structure transition observed between 295 K and 2.5 K in $(Li_{0.8}Fe_{0.2})OHFeSe$.

Fig. 3(a) shows the temperature dependence of the magnetic susceptibility $\chi$ with an external magnetic field of 10 Oe. The as-synthesized sample shows a round diamagnetic transition around 40 K, and a considerable shielding fraction of 60% at 2 K in the zero-field-cooling process, indicating bulk superconductivity. In order to identify the carrier type in $(Li_{0.8}Fe_{0.2})OHFeSe$, temperature dependent Seebeck coefficient was measured as shown in Fig. 3(b). The Seebeck coefficient is always negative in the whole temperature range from 4.5 to 300 K, indicating that electron-type carriers dominate in $(Li_{0.8}Fe_{0.2})OHFeSe$. The absolute value of thermoelectric power, │S│, is 6.27 μV/K at 300 K, and increases with decreasing temperature, and reaches maximum of 60 μV/K at ~140 K. Such a domelike behavior has been observed in other iron-based superconductors including $LnFe_{1-x}Co_xAsO$[26], $\beta$-$FeSe$[27] and $K_xFe_{2-y}Se_2$[28]. A rapid decrease in the absolute value of thermoelectric power takes place at 40 K and finally the thermoelectric power reaches zero due to superconducting transition, consistent with that observed in the magnetic susceptibility.

The magnetic susceptibility ($\chi$) as a function of temperature from 2 K to 300 K under an external field of 1T for the as-synthesized samples is shown in Fig. 4(a). The superconductivity is completely suppressed under this field. The temperature dependence of magnetization displays a Curie-Weiss behavior above 10 K, while a sudden decrease occurs in the ZFC curve around 8.5 K. The FC and ZFC magnetic susceptibilities bifurcate around 8.5 K. It suggests a weak ferromagnetic component due to a canted antiferromagnetic order, as derived from nuclear magnetic resonance (NMR) measurements. In order to confirm the magnetic transition, thermodynamic measurements were performed. Fig. 4(b) shows the specific heat under different magnetic fields. Specific heat begins to rise at about 8.5 K, which is consistent with the transition temperature in the magnetic susceptibility. Such rise is suppressed with increasing magnetic fields and becomes very obscure as the field increases up to 9 T. NMR experiment indicates that the canted antiferromagnetic (AFM) order originates from (LiFe)OH layer and the AFM could coexist with superconductivity. However, this antiferromagnetic order could not be observed in the NPD experiment because of the small magnetic moment. High quality single crystal is thus needed to reveal the possible magnetic order by neutron scattering measurements.

We successfully synthesize a new FeSe-derived superconductor $(Li_{0.8}Fe_{0.2})OHFeSe$ with $T_c$ of 40 K by a novel hydrothermal synthesis method. Via the combination of powder X-ray diffraction, neutron powder diffraction and nuclear magnetic resonance, the structure of $(Li_{0.8}Fe_{0.2})OHFeSe$ is unambiguously determined. With alternate stacking of $(Li_{0.8}Fe_{0.2})OH$ layer and anti-PbO type FeSe layer, there exists weak hydrogen bonding interaction between the layers. Compared to $\beta\text{-}Fe_{1+\delta}Se$, the $FeSe_4$ tetrahedron is extremely compressed in the *ab* plane in $(Li_{0.8}Fe_{0.2})OHFeSe$. Such distorted $FeSe_4$ tetrahedron is likely to play a key structural role in enhancing the superconductivity. In addition, Seebeck coefficient measurement reveals that electron-type carriers dominate in this material, in accordance with the refined composition which indicates a charge transfer from $(Li_{0.8}Fe_{0.2})OH$ layer to FeSe layer. Susceptibility, specific heat and NMR measurements indicate that a canted antiferromagnetic order occurs at ~8.5 K, and coexists with superconductivity at 40 K. Different from other FeSe-derived superconductors which are extremely air-sensitive, $(Li_{0.8}Fe_{0.2})OHFeSe$ superconductor is air-stable, and high quality single crystal is expected to be very useful for studying the underlying physics of high-$T_c$ iron-based superconductors.

**Methods**

**Sample Synthesis:** Polycrystalline (Li$_{0.8}$Fe$_{0.2}$)OHFeSe samples were prepared by the hydrothermal reaction method. 0.012–0.02 mol selenourea (Alfa Aesar,[29] 99.97% purity), 0.0075 mol Fe powder (Aladdin Industrial, Analytical reagent purity), and 12 g LiOH·H$_2$O (Sinopharm Chemical Reagent, A.R. purity) were put into a Teflon-lined steel autoclave (50 mL) and mixed together with 10 ml de-ionized water. The Teflon-lined autoclave was then tightly sealed and heated at 160 ℃ for 3–10 days. The polycrystalline samples acquired from the reaction system were shiny lamellar, which were then washed repeatedly with de-ionized water and dried at room temperature.

**Structure characterization:** XRD data were collected by using X-ray diffractometer (SmartLab-9, Rikagu Corp.) with Cu K$_\alpha$ radiation and a fixed graphite monochrometer with the 2-theta range of 5-80 ° and a scanning rate of 0.1 °/min at room temperature. NPD experiments were conducted using Ge311 (λ=2.0775Å) and Cu311 (λ=1.5401Å) monochrometer, respectively. Data were collected over the 2-Theta range of 1.3-166.3 °(Ge311) and 3-168 °(Cu311) with a step size of 0.05 ° from 2.5 K to 295 K. The NPD experiments were carried out in NIST Center for Neutron Research.

**Magnetic susceptibility and thermoelectric power:** Magnetization measurements were carried out by SQUID MPMS-XL5 (Quantum Design). The thermoelectric power was measured using a laboratory made system with a differential method.

**Nuclear magnetic resonance:** Standard NMR spin-echo techniques were used with commercial NMR spectrometer from Thamway Co. Ltd. The external magnetic field was generated by a 12 Tesla magnet from Oxford Instruments. Fine powder samples were used and packed in a quartz tube before put into NMR coils made by copper or silver. The $^{27}$Al NMR signal from 0.8 μm-thick aluminum foils (99.1 % purity) was used to calibrate the external field. Both $^{7}$Li and $^{1}$H NMR spectra were obtained by sweeping the frequency at fixed magnetic field values and integrating the spin-echo signal for each frequency value. Spin-spin relaxation time (T$_2$) measurement was carried out at fixed frequency at all temperatures and the spin-echo decay was fitted by exponential decay formula $I(t)=I_b+I_0\exp(-t/T_2)$.

**Acknowledgements**

This work is supported by National Natural Science Foundation of China (NSFC), the "Strategic Priority Research Program (B)" of the Chinese Academy of Sciences, the National Basic Research Program of China (973 Program), and Chinese Academy of Sciences.

**Author contributions**

X.F.L. and N.Z.W. contribute equally to this work. X.F.L. and N.Z.W. performed sample synthesis and susceptibility, specific heat, X-ray diffraction and thermoelectric power measurements with assistant from X.Z.Z. and X.G.L. Q.Z.H. H.W. and W.B. performed NPD experiment and did the structure analysis. Y.P.W., Z.D. and T.W. performed NMR experiment and analyzed data. G.H.Z. and F.Q.H. did the refinement on XRD. X.F.L., N.Z.W., Q.Z.H., T.W. and X.H.C. analyzed the data and wrote the paper. X.H.C. conceived and coordinated the project, and is responsible for the infrastructure and project direction. All authors discussed the results and commented on the manuscript.

**Additional information**

The authors declare no competing financial interests. Correspondence and requests for materials should be addressed to chenxh@ustc.edu.cn.

TABLE I. The crystallographic parameters from the NPD refinement of $(Li_{0.8}Fe_{0.2})OHFeSe$ at 2.5K. Space group: $P4/nmm$(No. 129); $a=b=3.77871(4)$ Å, $c=9.1604(1)$ Å, $V=130.7981$ Å$^3$, $R_{wp}=0.0183$, $R_p=0.015$, $\chi^2=1.498$. Due to the total neutron scattering amplitude for the $2a$ site is close to zero, the temperature factor for this site was fixed at 0.8.

| Atom | Wyckoff site | x | y | z | Occup | Uiso(x100 Å$^2$) |
|---|---|---|---|---|---|---|
| H | 2c | 0.75 | 0.75 | 0.1734(3) | 1.00 | 2.79(8) |
| O | 2c | 0.25 | 0.25 | -0.0764(1) | 1.00 | 0.27(4) |
| Li/Fe1 | 2a | 0.75 | 0.25 | 0 | 0.81/0.19(1) | 0.8 |
| Fe2 | 2b | 0.75 | 0.25 | 0.5 | 1.00 | 0.52(2) |
| Se | 2c | 0.25 | 0.25 | 0.3390(1) | 1.00 | 0.35(3) |

Interatomic distances (Å) and angles (degree)

| | |
|---|---|
| Fe2-Se | 2.4026(8) ×4 |
| Fe1/Li-O | 2.0149(6) ×4 |
| H-Se | 3.068(2) ×4 |
| O-H | 0.889(5) |
| | |
| Se$^{top}$-Fe2-Se$^{top}$ | 103.70(5) ×2 |
| Se$^{top}$-Fe2-Se$^{bottom}$ | 112.43(3) ×4 |
| O$^{top}$-Li/Fe1-O$^{top}$ | 139.3(1) ×2 |
| O$^{top}$-Li/Fe1-O$^{bottom}$ | 96.94(3) ×4 |
| O-H-Se | 119.43(6) ×4 |

# Figures:

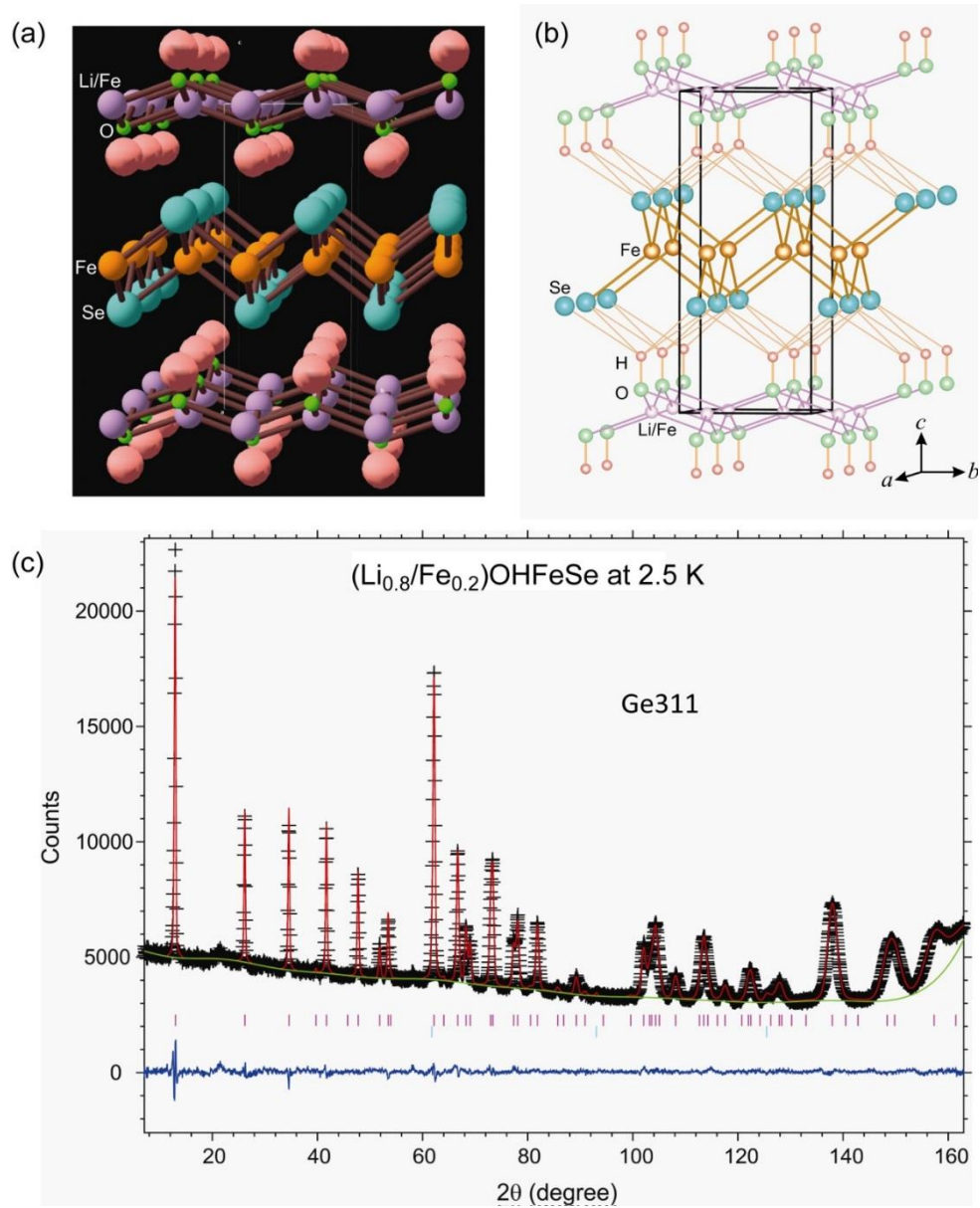

**Figure 1: (a)**: Neutron scattering Fourier difference analysis using the 2.5 K NPD data. Fourier difference contours are highlighted in pink. **(b)**: A schematic view of the structure of $(Li_{0.8}Fe_{0.2})OHFeSe$. In the model the anti-PbO-type FeSe layers and the $(Li_{0.8}Fe_{0.2})OH$ layers stack alternately. **(c)**: Observed (crosses) and calculated (red solid line) NPD pattern for $(Li_{0.8}Fe_{0.2})OHFeSe$ ($\lambda=2.0775$Å) at the 2.5 K. Differences between the observed and calculated intensities are shown in bottom of the figure. Bragg peak positions were indicated by short purple vertical bars below the NPD patterns.

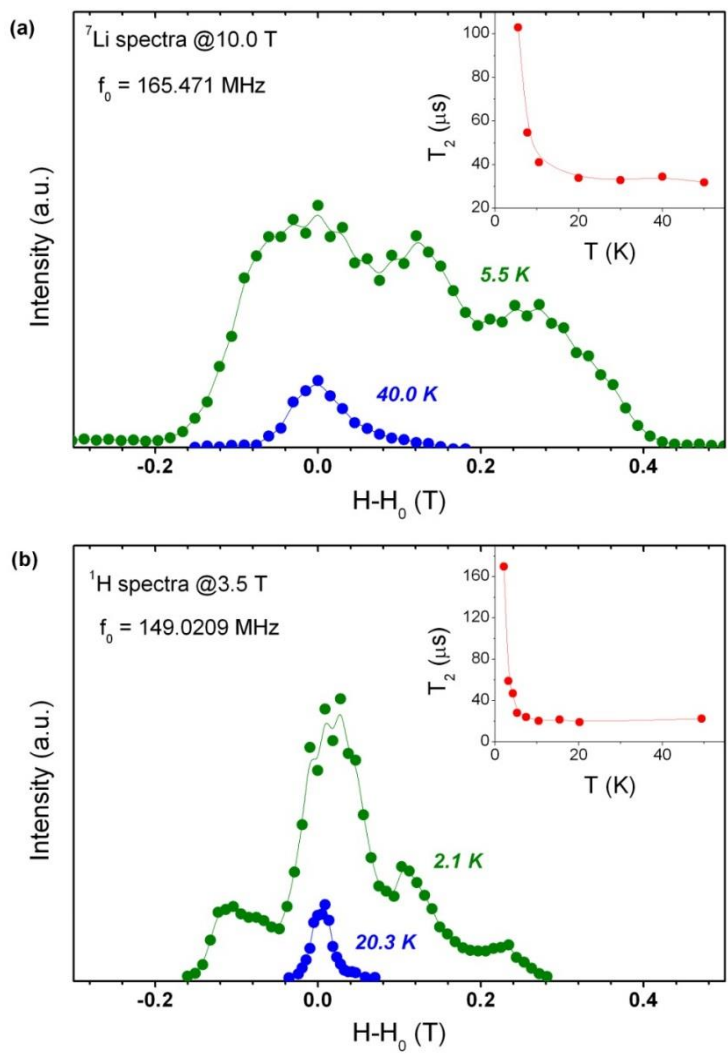

**Figure 2**: Evidence for Li and H in the same structure from NMR experiment. **(a)**: Temperature dependent [7]Li NMR spectra under external magnetic field H = 10.0 Tesla and the inset plot shows temperature dependent $T_2$ with NMR frequency $f_0$ = 165.471 MHz. **(b):** Temperature dependent [1]H NMR spectra under external magnetic field H = 10.0 Tesla and the inset plot shows temperature dependent $T_2$ with NMR frequency $f_0$ = 149.0209 MHz.

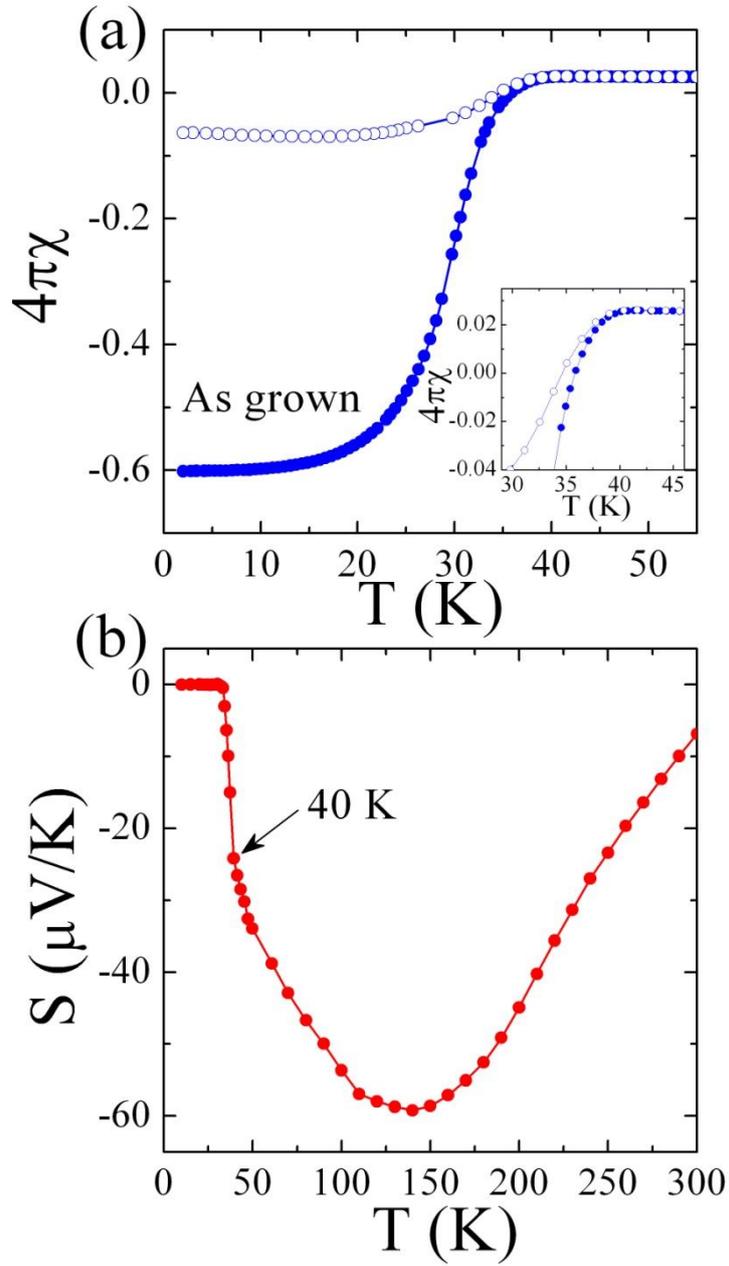

**Figure 3**: **(a):** The temperature dependence of magnetic susceptibility χ for the as-synthesized ($Li_{0.8}Fe_{0.2}$)OHFeSe sample. Data were collected under a magnetic field of 10 Oe. The inset in (a) is the zoom-in view around the superconducting transition of the as-synthesized sample. **(b):** The temperature dependent thermoelectric power of the as-synthesized ($Li_{0.8}Fe_{0.2}$)OHFeSe sample.

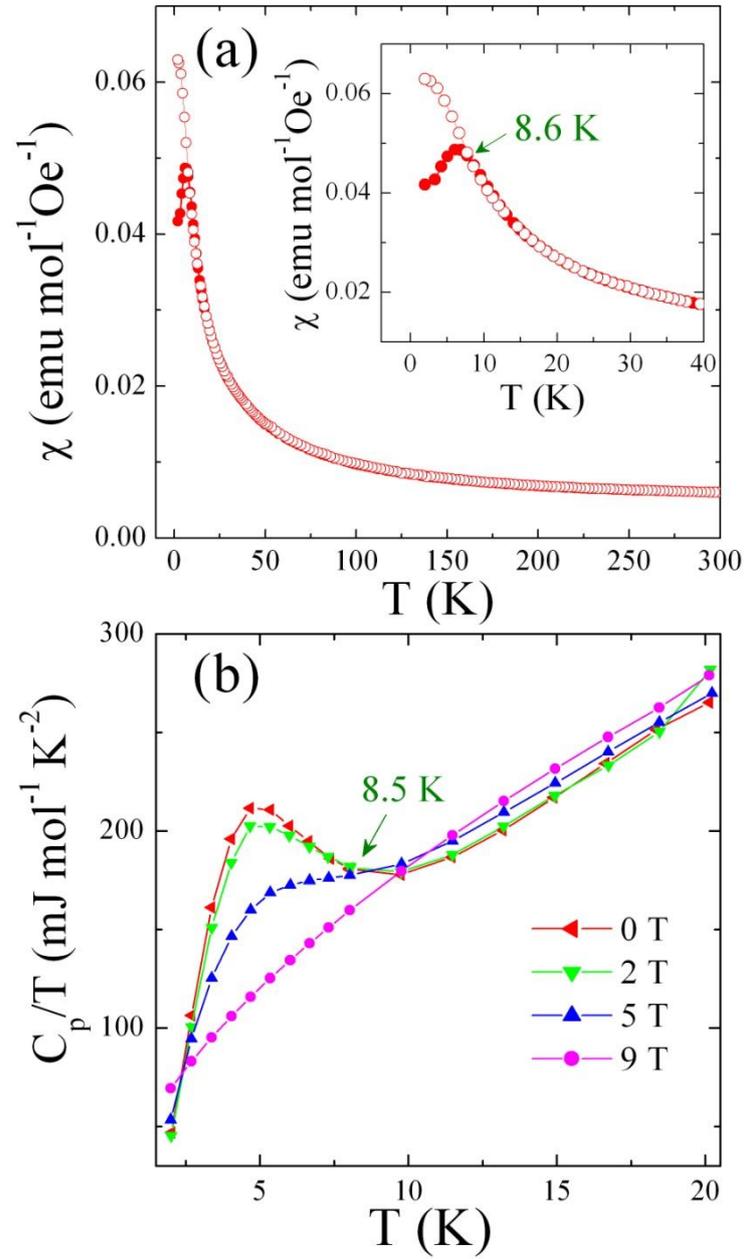

**Figure 4**: **(a):** Magnetic susceptibility ($\chi$) as a function of temperature for the as-synthesized $(Li_{0.8}Fe_{0.2})OHFeSe$ sample from 2-300 K with an external field of 1 T. **(b):** The specific heat of $(Li_{0.8}Fe_{0.2})OHFeSe$ under different external fields.

# Supplementary information

As is shown in Fig S1(a) and Fig S2(a), the Rietveld refinement using X-ray powder diffraction data reveals the crystal structure is composed of alternating anti-PbO FeSe layer and anti-PbO type $LiFeO_2$ layer. The initial neutron powder diffraction indicated that there is no Li/Fe existing in this structure while only H-O connected with FeSe layer present in this compound as shown in Fig S1(b) and Fig S2(b). However, this structure proposed by NPD is extremely different from that determined by X-ray diffraction experiment. Therefore, the nuclear magnetic resonance was conducted to understand the different preliminary structural models suggested by X-ray and neutron diffraction experiments. NMR result definitely proved that both of Li and H should be contained in the same structure. After considering the NMR result with both Li and H in the structure, the final structure of $(Li_{0.8}Fe_{0.2})OHFeSe$ is proposed and determined via Rietveld refinement of X-ray and neutron powder diffraction and NMR analysis. The $(Li_{0.8}Fe_{0.2})OHFeSe$ structure and the corresponding Rietveld refinement pattern of NPD data are shown in Fig S1(c) and Fig S2(c), respectively. In this structure, FeSe layer is the conducting block, while $(Li_{0.8}Fe_{0.2})OH$ layer is the charge reservoir block.

The refined structure parameters of $(Li_{0.8}Fe_{0.2})OHFeSe$ from NPD data at different temperatures from 2.5 K to 295 K are summarized in Table S1. Selected bond distances and bond angles for the final $(Li_{0.8}Fe_{0.2})OHFeSe$ structure model at different temperatures are also listed in Table S2.

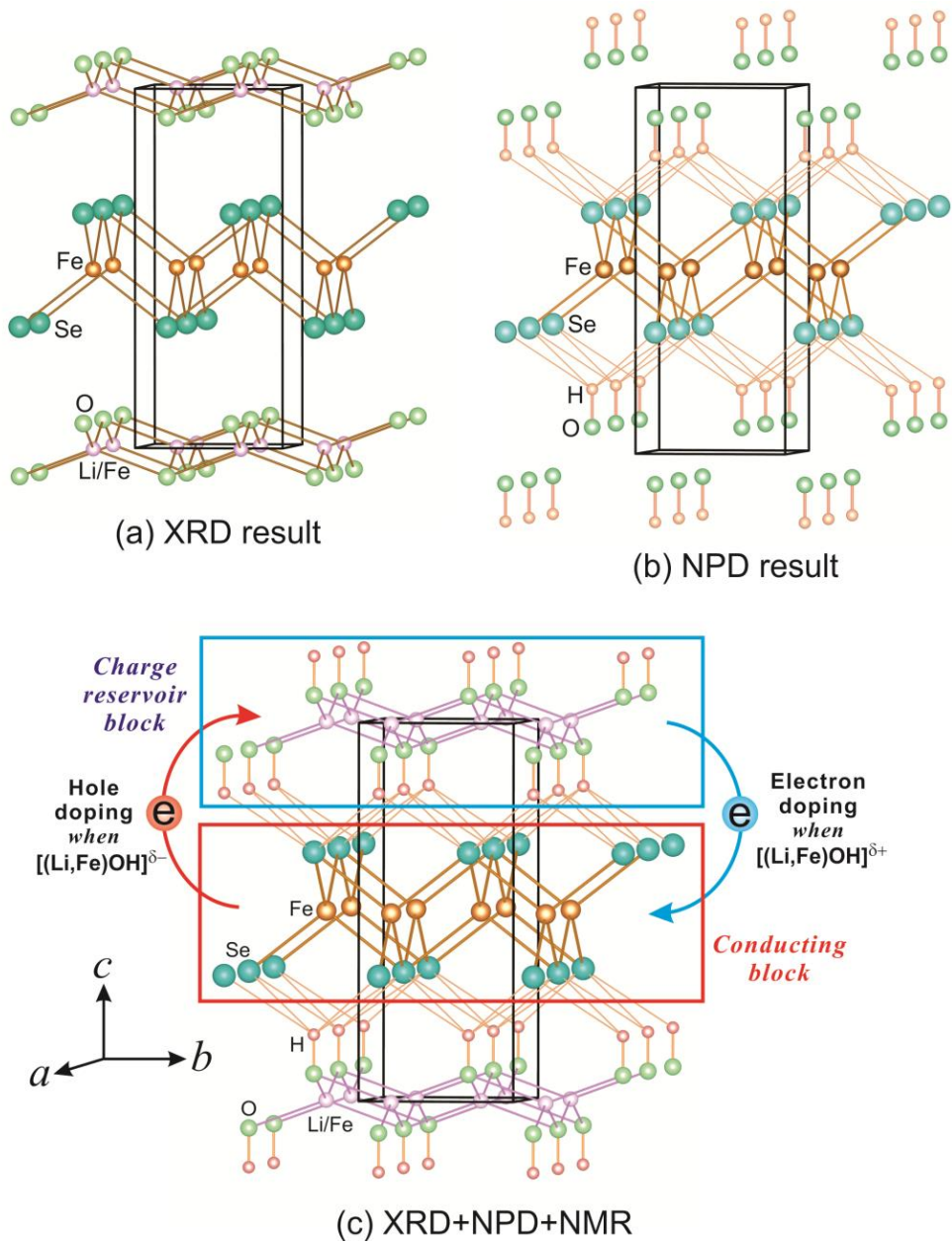

**Figure S1**. **(a):** Crystal structure of LiFeO$_2$Fe$_2$Se$_2$ model, which was determined from powder X-ray diffraction data. **(b):** A schematic structure of (OH)FeSe model from neutron powder diffraction data refinement result. In this model, OH layer and the FeSe layer are alternately stacked. **(c):** A schematic view of the structure of (Li$_{0.8}$Fe$_{0.2}$)OHFeSe, which was determined by the combination of powder X-ray diffraction, neutron powder diffraction and nuclear magnetic resonance. The FeSe layer, which acts as the conducting block, alternately stacks with the charge reservoir block (Li$_{0.8}$Fe$_{0.2}$)OH layer in this model.

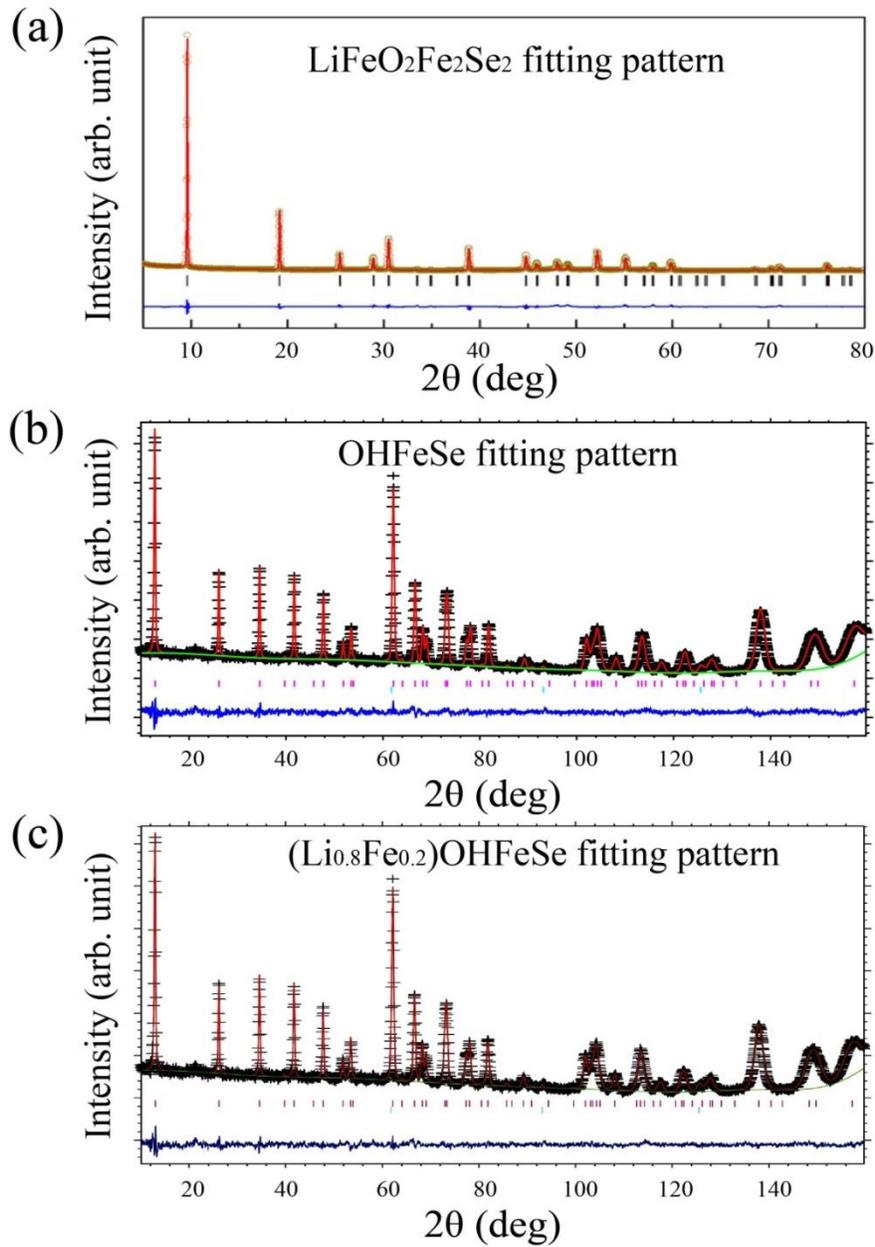

**Figure S2**: **(a):** Rietveld refinement on XRD data collected at 298 K for the structural model of LiFeO$_2$Fe$_2$Se$_2$ which was shown in Fig. S1(a). **(b):** Rietveld refinement on NPD data with the structural model of OHFeSe shown in Fig. S1(b). The NPD data was collected over the 2-Theta range of 1.3-166.3 ° using Ge311 (λ=2.0775Å) monochrometer at 4 K. **(c):** Rietveld refinement on NPD pattern which was collected at 4 K using Ge311 (λ=2.0775Å) monochrometer with the structural model of (Li$_{0.8}$Fe$_{0.2}$)OHFeSe shown in Fig. S1(c).

Table S1 Structural parameters of $(Li_{0.8}Fe_{0.2})OHFeSe$ from the NPD refinement for different temperature from 2.5 K to 295 K.

**2.5 K:** $a$=3.77871(4) Å and c=9.1604(1) Å,  Rwp=0.0183, Rp=0.015, $\chi^2$=1.498

| Atom | Wyckoff Site | x | y | z | Occup. | Uiso(x100 Å$^2$) |
|---|---|---|---|---|---|---|
| H | 2c | 0.75 | 0.75 | 0.1734(3) | 1.00 | 2.79(8) |
| O | 2c | 0.25 | 0.25 | -0.0764(1) | 1.00 | 0.27(4) |
| Li | 2a | 0.75 | 0.25 | 0 | 0.81(1) | 0.8 (fixed) |
| Fe1 | 2a | 0.75 | 0.25 | 0 | 0.19(1) | 0.8 (fixed) |
| Fe2 | 2b | 0.75 | 0.25 | 0.5 | 1.00 | 0.52(2) |
| Se | 2c | 0.25 | 0.25 | 0.3390(1) | 1.00 | 0.35(3) |

**4 K:** $a$=3.77887(6) Å and c=9.1609(2) Å, Rwp=0.0309, Rp=0.0251, $\chi^2$=1.141

| Atom | Wyckoff Site | x | y | z | Occup. | Uiso(x100 Å$^2$) |
|---|---|---|---|---|---|---|
| H | 2c | 0.75 | 0.75 | 0.1765(4) | 1.00 | 3.08(8) |
| O | 2c | 0.25 | 0.25 | -0.0759(2) | 1.00 | 0.59(4) |
| Li | 2a | 0.75 | 0.25 | 0 | 0.81(1) | 0.8 (fixed) |
| Fe1 | 2a | 0.75 | 0.25 | 0 | 0.19(1) | 0.8 (fixed) |
| Fe2 | 2b | 0.75 | 0.25 | 0.5 | 1.00 | 0.55(2) |
| Se | 2c | 0.25 | 0.25 | 0.3386(1) | 1.00 | 0.29(3) |

**15 K:** $a$=3.77947(7) Å and c=9.1617(3) Å, Rwp=0.0275, Rp=0.0223, $\chi^2$=1.139

| Atom | Wyckoff Site | x | y | z | Occup. | Uiso(x100 Å$^2$) |
|---|---|---|---|---|---|---|
| H | 2c | 0.75 | 0.75 | 0.1751(5) | 1.00 | 3.3(1) |
| O | 2c | 0.25 | 0.25 | -0.0748(2) | 1.00 | 0.70(4) |
| Li | 2a | 0.75 | 0.25 | 0 | 0.80(1) | 0.8 (fixed) |
| Fe1 | 2a | 0.75 | 0.25 | 0 | 0.20(1) | 0.8 (fixed) |
| Fe2 | 2b | 0.75 | 0.25 | 0.5 | 1.00 | 0.46(4) |
| Se | 2c | 0.25 | 0.25 | 0.3369(2) | 1.00 | 0.54(3) |

**295 K:** $a$=3.7860(1) Å and c=9.2880(9) Å, Rwp=0.0499, Rp=0.0412, $\chi^2$=0.9065

| Atom | Wyckoff Site | x | y | z | Occup. | Uiso(x100 Å$^2$) |
|---|---|---|---|---|---|---|
| H | 2c | 0.75 | 0.75 | 0.174(1) | 1.00 | 5.0(2) |
| O | 2c | 0.25 | 0.25 | -0.0737(6) | 1.00 | 2.2(1) |
| Li | 2a | 0.75 | 0.25 | 0 | 0.82(1) | 1.5 (fixed) |
| Fe1 | 2a | 0.75 | 0.25 | 0 | 0.18(1) | 1.5 (fixed) |
| Fe2 | 2b | 0.75 | 0.25 | 0.5 | 1.00 | 1.57(5) |
| Se | 2c | 0.25 | 0.25 | 0.3384(5) | 1.00 | 1.70(7) |

Table S2 Selected bond distance (Å) and bond angles (°) of $(Li_{0.8}Fe_{0.2})OHFeSe$ for different temperature from 2.5K to 295K.

| | | 2.5K | 4K | 15K | 295K |
|---|---|---|---|---|---|
| Fe1/Li-O | ×4 | 2.0149(6) | 2.0134(7) | 2.0103(9) | 2.013(2) |
| Fe2-Se | ×4 | 2.4026(8) | 2.3990(8) | 2.4091(12) | 2.416(3) |
| H-Se | ×4 | 3.068(2) | 3.057(2) | 3.056(3) | 3.078(7) |
| O-H | | 0.889(5) | 0.922(5) | 0.919(7) | 0.94(2) |
| | | | | | |
| O-Li/Fe1-O | ×2 | 139.3(1) | 139.6(1) | 140.1(1) | 140.2(3) |
| | ×4 | 96.94(3) | 96.85(3) | 96.68(5) | 96.7(1) |
| H-O-Li/Fe1 | ×4 | 110.34(5) | 110.21(5) | 109.94(7) | 109.9(2) |
| Se-Fe2-Se | ×2 | 103.70(5) | 103.92(5) | 103.33(7) | 103.2(2) |
| | ×4 | 112.43(3) | 112.313 | 112.63(4) | 112.70(9) |
| O-H-Se | ×4 | 119.43(6) | 119.06(7) | 119.01(9) | 119.7(2) |
| Se-H-Se | ×4 | 76.03(5) | 76.35(6) | 76.39(8) | 75.9(2) |
| | ×2 | 121.2(1) | 121.9(1) | 122.0(2) | 120.9(5) |